# Highly Efficient Bilateral Doping of Single-Walled Carbon Nanotubes

*Anastasia E. Goldt[a#], Orysia T. Zaremba[a#], Mikhail O. Bulavskiy[a], Fedor S. Fedorov[a],*

*Konstantin V. Larionov[b,c], Alexey P. Tsapenko[d], Zakhar I. Popov[b,e], Pavel Sorokin[b],*

*Anton S. Anisimov[f], Heena Inani[g], Jani Kotakoski[g], Kimmo Mustonen[g], Albert G. Nasibulin[a,h*]*

[a]Skolkovo Institute of Science and Technology, 3 Nobel Street, 121205 Moscow, Russia

[b] National University of Science and Technology "MISiS", Leninsky prospect 4, 119049 Moscow, Russian Federation

[c]Moscow Institute of Physics and Technology, Institutskiy lane 9, Dolgoprudny, 141700 Moscow region, Russian Federation

[d]Aalto University, Department of Applied Physics, Puumiehenkuja 2, 00076 Espoo, Finland

[e]Emanuel Institute of Biochemical Physics RAS, 119334 Moscow, Russian Federation

[f]Canatu Ltd., Konalankuja 5, 00390 Helsinki, Finland

[g]University of Vienna, Faculty of Physics, Boltzmanngasse 5, 1090 Vienna, Austria

[h]Aalto University School of Chemical Engineering, Kemistintie 1, 16100 Espoo, Finland

ABSTRACT

A boost in the development of flexible and wearable electronics facilitates the design of new materials to be applied as transparent conducting films (TCFs). Although single-walled carbon nanotube (SWCNT) films are the most promising candidates for flexible TCFs, they still do not meet optoelectronic requirements demanded their successful industrial integration. In this study, we proposed and thoroughly investigated a new approach that comprises simultaneous bilateral (outer and inner surfaces) SWCNT doping after their opening by thermal treatment at 400 °C under an ambient air atmosphere. Doping by a chloroauric acid ($HAuCl_4$) ethanol solution allowed us to achieve the record value of sheet resistance of 31 ± 4 Ω/sq at a transmittance of 90% in the middle of visible spectra (550 nm). The strong *p*-doping was examined by open-circuit potential (OCP) measurements and confirmed by *ab initio* calculations demonstrating a downshift of Fermi level around 1 eV for the case of bilateral doping.

INTRODUCTION

Transparent conducting films are now in the limelight of modern electronics driven by the expanding field of flexible devices that include displays, touch screens, solar cells, etc.[1–3] Besides high transparency and good conductivity, such applications require rather good flexibility of materials to be preferably made under cost-effective fabrication protocols. Commonly applied transparent metal oxides, such as indium-tin-oxide (ITO), have several drawbacks, including high refractive index, limited flexibility, restricted chemical robustness, and depleted raw material supply. As a result, ITO is being impugned by alternatives like Ag nanowires mesh, PEDOT:PSS,[5] graphene,[6] reduced graphene oxide,[7] and single-walled carbon nanotubes (SWCNTs).[8]

At the forefront, the materials such as SWCNTs are believed to be advantageous candidates for the ITO replacement due to their excellent optoelectronic properties, chemical stability, an abundant amount of carbon, and good adhesion to various substrates.[9–12] Their applications enable us to create entirely new components in flexible and stretchable transparent electronics.[13–15] However, in addition to carbon nanotube parameters such as length, the concentration of defects, degree of bundling, their optoelectronic properties depend greatly on the ratio of semiconducting and metallic tubes, which directly impacts CNT-CNT junctions.

The most radical and efficient way to improve both optical and electrical properties of SWCNTs can be realized through tuning their Fermi level by adsorption doping.[16,17] This approach employs *p*- or *n*-doping compounds, which cause a shift of electron density depending on their redox potential relative to the potential of SWCNTs when adsorbed on their surfaces.[18] For redox species, more positive electrode potentials than that of the SWCNTs yield the efficient Fermi level downshift (*p*-doping) resulting in (i) transition of contacts between metallic and semiconducting SWCNTs from Schottky to Ohmic, (ii) increase in charge carriers concentration, which in turn leads to decrease in overall resistance, (iii) suppression of adsorption peaks associated with the transition between van Hove singularities (vHs), *i.e.*, leading to the improvement in transmittance. Among various dopants, chloroauric acid ($HAuCl_4$) has been found to be the most efficient one that enabled to achieve the state-of-the-art sheet resistance of 39 Ω/sq with the transmittance of 90% at 550 nm.[19]

An appealing strategy to dope SWCNTs comprises the nanotube opening and subsequent dopant encapsulation that protects a dopant from ambient conditions. The most commonly used methods for SWCNT filling are melting or evaporation of the desired compound in a vacuumed ampoule in the presence of nanotubes. Although these approaches were successfully applied for a wide range of metals, salts, and other materials, they are time-consuming, require vacuum (~$10^{-5}$ Torr) and high-temperature conditions (100 - 1000 °C), lack efficiency, and exhibit low yield.[20–22]

Here, we propose a simple method to efficiently improve the optoelectronic properties, comprising thermal treatment of SWCNT films at an environmental atmosphere and subsequent doping by HAuCl$_4$ solution to deliver the dopant to the outer surface and inner space of SWCNTs. Such a strategy has allowed us to achieve a record value of the equivalent sheet resistance of 31 Ω/sq with the transmittance of 90% (at 550 nm), which stems from the bilateral nanotube doping as confirmed by density functional theory (DFT) calculations and electrochemical studies.

RESULTS AND DISCUSSION

DOPING PROCEDURE

For high-efficiency SWCNT doping, we implemented a facile method of thermal treatment in an ambient atmosphere with a subsequent drop-casting of HAuCl$_4$ solution. Thermal treatment of SWCNT films in the range of 300 - 500 °C facilitates opening the tubes, *i.e.* by the cap removal (as discussed in detail in Supporting information), exposing the inner surface of the tubes to the environment. To increase conductivity and transmittance of the SWCNT networks, we utilized one of the strongest electron acceptor dopants for SWCNT films – chloroauric acid (HAuCl$_4$).[23] Extraction of electron density from SWCNTs promotes the reduction of gold from oxidized state to metallic one, which decorates SWCNT surface in the form of Au$^0$ nanoparticles.[24]

Figure 1 shows the equivalent sheet resistance and transmittance as a function of the thermal treatment temperature of pristine SWCNTs doped with 15 mM HAuCl$_4$ ethanol solution. The equivalent sheet resistance, $R_{90}$ is a quantitative measure for conductive transparent materials that balances the sheet resistance and transmittance of SWCNT films. The equivalent sheet resistance is usually assessed at the film transmittance of $T = 90\%$ as $R_{90}=1/[K \log(10/9)]$ with the quality factor of $K=1/(R_S \cdot A)$, where $A$ is an absorbance measured at 550 nm and $R_S$ denotes the sheet resistance (Ω/sq).[25] While the pristine SWCNTs typically had equivalent sheet resistance of about 150 Ω/sq, drop-casting of the HAuCl$_4$ solution yielded a reduction of $R_{90}$ by 65%. After the treatment at 300 - 400 °C, doped SWCNTs exhibit the lowest $R_{90}$ value approaching a record of 31 ± 4 Ω/sq (at 400 °C). A change in $R_{90}$ is primarily attributed to *p*-doping by [AuCl$_4$]$^-$ ion driven by its high standard electrochemical potential, [18,26] which is high enough to force the first and the second vHs electrons to disappear, especially pronounced in the case of tubes with a large diameter.[24] Such doping effect is manifested at UV-vis-NIR transmittance spectra (Figure 1b), leading to the removal of $S_{11}$, $S_{22}$, and $M_{11}$ peaks and the appearance of a new peak that corresponds to the intersubband plasmon of nanotubes appearing when excitonic levels are fully saturated with charge carriers.[27,28] These data prove that *p*-doping caused an efficient downshift of Fermi level favored by the HAuCl$_4$ treatment.

An increase in the transmittance after the doping of nanotubes treated at temperatures above 300 °C (Figure S1) can be attributed to the entering of the dopant into the SWCNTs' inner space, favored by the opening of the tubes. At temperatures above 400 °C, we observe a deterioration of the sheet resistance that can be explained by the formation of numerous defects due to the destructive oxidation of SWCNTs. Although at 400 °C higher transmittance should have impacted $R_{90}$, this effect is negated by a huge sheet resistance jump.

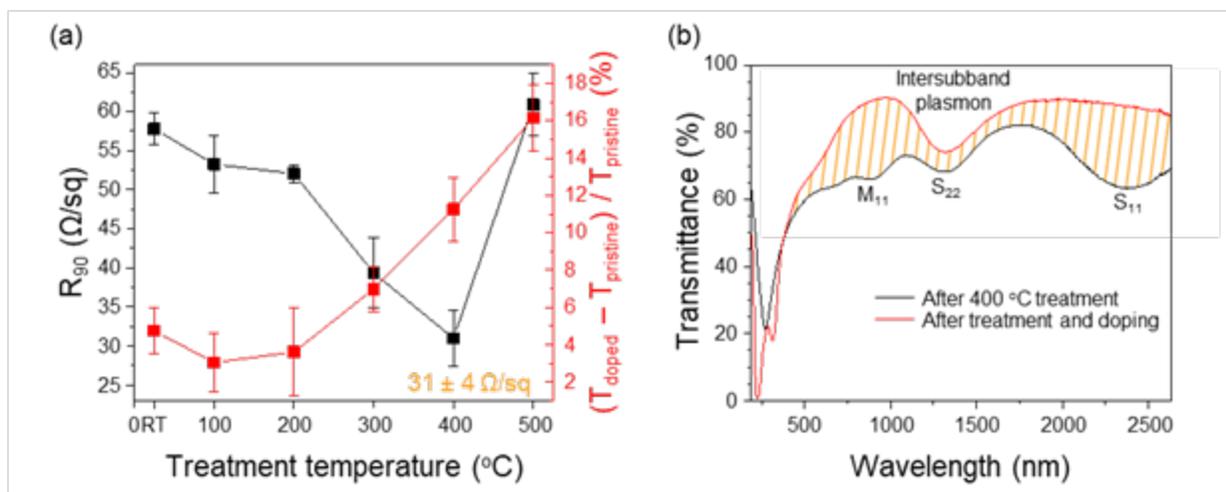

**Figure 1.** Effect of thermal treatment and doping on optoelectronic properties of SWCNTs. (a) Equivalent sheet resistance and transmittance of the SWCNT films treated at different temperatures (100 - 500 °C) for 20 min followed by doping with 15 mM HAuCl$_4$ ethanol solution; (b) UV-vis-NIR absorption spectra of SWCNT film treated at 400 °C in the air (black) and the same film after doping with 15 mM HAuCl$_4$ (red); hatched area illustrates film transmittance increase in the visible and NIR range due to the doping.

CHARACTERIZATION OF DOPED SWCNTs

We have evaluated the doping efficiency of SWCNT films treated at 100 - 500 °C using Raman spectroscopy. Raman spectrum of the pristine SWCNT films (Figure 2a) typically contains chirality sensitive radial breathing mode (RBM) around 100 - 500 cm$^{-1}$, D-mode peak near 1340 cm$^{-1}$, graphitic G-mode peak near 1590 cm$^{-1}$, and 2D-mode peak at 2680 cm$^{-1}$. The RBM is an out-of-plane bond-stretching mode, where peak wavenumbers are inversely proportional to the SWCNT diameter, D-mode reflects the concentration of structural defects, G-mode is responsible for tangential vibrations of carbon atoms, and 2D-mode is the second harmonic overtone of the D-mode.[29]

Figures 2b,c present respectively G-mode and 2D-mode changes of SWCNT films treated at 100 - 500 °C with subsequent $HAuCl_4$-doping. Typically, blue shifts of the G-mode peak and 2D mode positions are attributed to the *p*-doping. [17, 30] The G-mode peak of all samples shifts to higher values of *ca*. 25 - 28 cm$^{-1}$, while the G-mode peak of films treated at 400 and 500 °C demonstrates a slightly greater shift (29 cm$^{-1}$) when compared to the others (Figure 2b). 2D-mode peak intensities experience a remarkable decrease accompanied by peak position upshift. Similar to G-mode changes, 2D-mode alterations are most pronounced for 400 and 500 °C-treated SWCNT films (Figure 2c). Therefore, SWCNTs thermally treated at 100 - 500 °C experience strong *p*-doping. While the doping efficiency of films treated at 400 - 500 °C is higher than in the case of other temperatures, which can be attributed to the caps' removal at temperatures above 300 °C and effective doping of the opened SWCNTs.

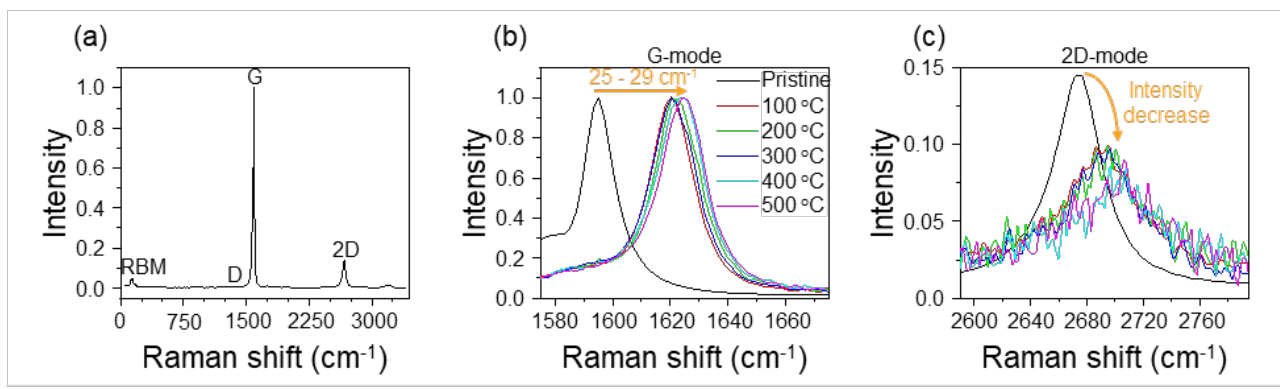

**Figure 2.** Raman spectra of pristine SWCNTs (a); G-mode (b) and 2D-mode (c) peaks of doped SWCNTs after thermal treatment at 100 - 500 °C in comparison to the pristine film.

The origin of more efficient SWCNT doping due to the prior caps' removal via thermal treatment is also supported by our transmission electron microscopy (TEM) observations. Figure 3 displays TEM images of $HAuCl_4$-doped films of both pristine and thermally treated SWCNTs. In images of pristine and treated up to 300 °C SWCNTs, we observe the decoration of nanotubes' surface with metallic gold nanoparticles with a size of approximately 5 - 10 nm (Figure 3a,b). In contrast, for samples thermally treated at 400 °C, in addition to outer surface decoration with $Au^0$ nanoparticles, we can notice the filling of the inner SWCNT space (Figure 3c,d). Measurement of the interplanar spacing of the encapsulated material from TEM images gave the result of 0.235 nm. It was admitted to be metallic gold, which has (111) interplanar spacing of 0.2355 nm (PDF-2 [4-784]). Gold nanoparticles decorate the outer surface of SWCNTs, formed via spontaneous reduction of $[AuCl_4]^-$ anions. When the heat-treatment temperature is high enough to oxidize the nanotube caps, *e.g.* 400 °C, doping solution penetrates the SWCNT inner space, resulting in improved doping and, also, the formation of the metallic gold phase. This is manifested in higher doping efficiency for SWCNT film treated at 400 °C, which explains its record $R_{90}$ value.

Although we attributed the formation of $Au^0$ nanoparticles and encapsulated gold wires to spontaneous reduction of $[AuCl_4]^-$, some parts of metallic gold can be formed under the exposure of 100 kV TEM electron beam[31]. Therefore, we carried out scanning transmission electron microscopy (STEM) measurements at a low voltage of 60 keV electron energy (Figure 3c,d and Figure S7). No transformation of nanowires inside nanotubes during the measurements was observed.

Figure 3e,f shows a gold wire at the edge of a nanotube bundle that precisely following the structure of the SWCNT and thus serves as strong evidence for their endohedral nature. During *in situ* measurements, the diffusion of gold atoms within the nanotube is observed; at 60 keV in a pseudo-elastic scattering from Au atoms electron energy reaches up to ~0.9 eV, which may overcome the cohesive energy at the disordered edge and release atoms. The diffusion of carbon and released Au atoms within the tube is observed, imparting a smeared-out contrast confined along the nanotube (Figure S7 and Video).

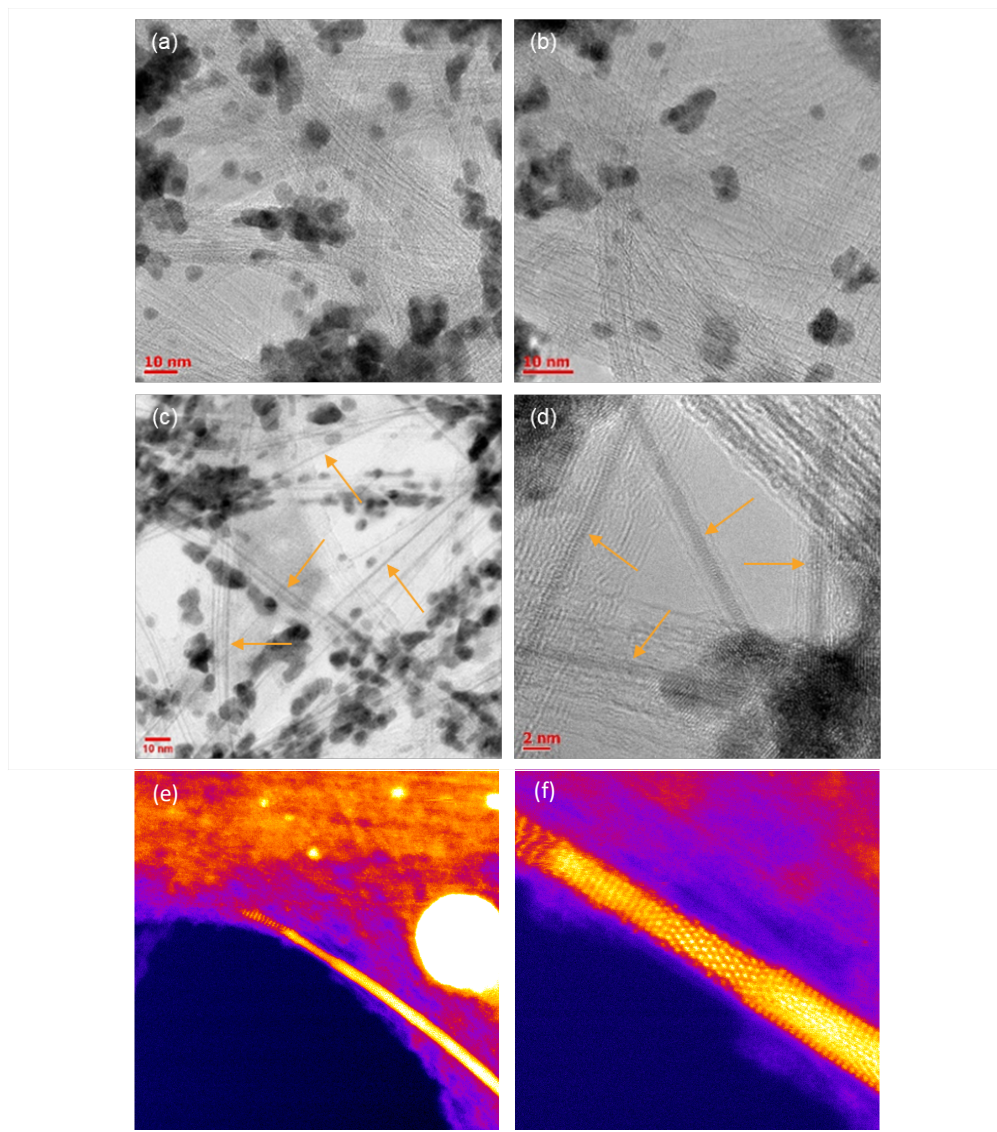

**Figure 3.** TEM images of SWCNT films doped with 15 mM HAuCl$_4$ ethanol solution: without preliminary thermal treatment (a); pre-treatment at 300 °C (b) and 400 °C (c,d). Arrows indicate SWCNTs filled with the metallic Au phase. STEM images of 15 mM HAuCl$_4$-doped opened SWCNTs with encapsulated gold nanowires are shown in (e,f). The respective field of views are 32 × 32 nm$^2$ and 8 × 8 nm$^2$.

EVALUATION OF THE DOPING MECHANISM

To explain the higher doping efficiency of treated nanotubes, we have evaluated changes by measuring OCP of SWCNT films immersed into the HAuCl$_4$ solution. A general trend related to OCP transients is illustrated in Figure 4a recorded for pristine and treated at 350 °C SWCNT films immersed into ethanol solutions with 7.5 and 60 mM HAuCl$_4$ concentrations. We observe that all OCP increase until they are stabilized after *ca*. 1600 s (Figure S5). These potential transients indicate the appearance of the electron transfer processes, i.e. the discharge process. In Figure 4b, steady-state potentials reached at 1600 s are presented as functions of dopant concentration and are fitted with the Nernst equation: $E = E^o + (R \cdot T/n \cdot F) \cdot \ln C_{[AuCl_4]^-}$. Hence, by extrapolating to 1 M concentration of HAuCl$_4$, we can assess the standard potentials ($E^o$) of corresponding doping processes, while a coefficient of the logarithm allows estimating the number of transferred electrons in the discharge process ($n$).

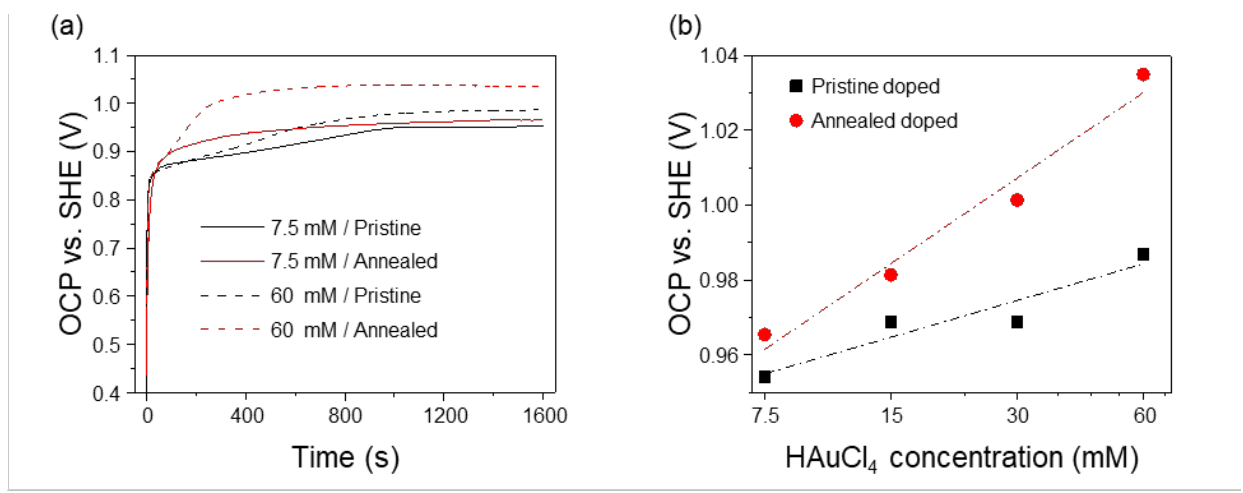

**Figure 4.** Open circuit potential (OCP) transients of pristine (black) and treated at 350 °C (red) SWCNTs recorded in 7.5 and 60 mM HAuCl$_4$ solutions (a); concentration dependence of OCP

values at 1600 s for 7.5, 15, 30 and 60 mM HAuCl$_4$ solutions (b). The potential scale is presented *versus* the standard hydrogen electrode (SHE).

From Figure 4a,b, we can observe that the equilibrium potential of the treated SWCNT film in comparison to the pristine ones becomes 11 and 48 mV greater for doping with 7.5 and 60 mM HAuCl$_4$ solutions, respectively. The more positive potential indicates the more efficient withdrawal of electrons from SWCNTs or more efficient doping.

When the SWCNT film is in contact with the HAuCl$_4$ ethanol solution, several charge-transfer involving half-reactions (Table 1) occur on the electrode.[24] From this set of equilibria, we see that monovalent gold chloride complexes coexist with trivalent gold chloride anions and metallic gold at the SWCNT films surface; thus these [AuCl$_2$]$^-$ ions can be formed via [AuCl$_4$]$^-$ reduction and Au$^0$ oxidation. Despite metastable behavior of [AuCl$_2$]$^-$ complexes in chloride solutions and their tendency of disproportionation at high concentrations,[26] in the studied system at the equilibrium ($E_{b1} = E_{b2} = E_{b3}$) ions of monovalent gold complexes exist with concentration of $C_{[AuCl_2]^-} = 2.68 \cdot 10^{-3} \cdot C_{[AuCl_4]^-}^{1/3} \cdot C_{Cl^-}^{2/3}$. If we compare standard electrode potential values of [AuCl$_4$]$^-$ and [AuCl$_2$]$^-$ reduction to metallic gold, we will see that the latter has greater $E^o$. It means that chloride complexes of monovalent gold have greater electron acceptors ability than trivalent gold ions and behave as even stronger *p*-dopant.

**Table 1.** Reactions occurring at the SWCNT film when it is in contact with the doping solution, their standard electrode potentials, and equations for electrode potentials.

|      | Half-reaction | Standard electrode potential | Electrode potential |
|------|---------------|------------------------------|---------------------|
| (a)  | $CNT \rightleftharpoons CNT^+ + 1e^-$ | $E_a^o = \pm\, 0.025\, V^{17,18}$ | |
| (b1) | $[AuCl_4]^- + 3e^- \rightleftharpoons Au^0 + 4Cl^-$ | $E_{b1}^o = 1.002\, V$ | $E_{b1} = E_{b1}^o + \dfrac{R \cdot T}{3 \cdot F} \cdot \ln \dfrac{C_{[AuCl_4]^-}}{C_{Cl^-}^4}$ |
| (b2) | $[AuCl_4]^- + 2e^- \rightleftharpoons [AuCl_2]^- + 2Cl^-$ | $E_{b2}^o = 0.926\, V$ | $E_{b2} = E_{b2}^o + \dfrac{R \cdot T}{2 \cdot F} \cdot \ln \dfrac{C_{[AuCl_4]^-}}{C_{[AuCl_2]^-} \cdot C_{Cl^-}^2}$ |
| (b3) | $[AuCl_2]^- + 1e^- \rightleftharpoons Au^0 + 2Cl^-$ | $E_{b3}^o = 1.154\, V$ | $E_{b3} = E_{b3}^o + \dfrac{R \cdot T}{1 \cdot F} \cdot \ln \dfrac{C_{[AuCl_2]^-}}{C_{Cl^-}^2}$ |

Table 2 shows standard electrode potential and transferred electrons values calculated from the fitting of $E - C_{[AuCl_4]^-}$ dependencies (Figure 4b) in comparison with $E^o$ and $n$ for $[AuCl_4]^-$ and $[AuCl_2]^-$ reduction to $Au^0$. The measured number of transferred electrons does not coincide with these of the referential processes because we do not consider the impact of Cl⁻ ions concentration on the potential for the fitting of equilibrated potential-dopant concentration dependencies. Indeed, such an impact should be contributive because chlorine anions concentration depends on the dopant concentration, although the functional dependence is rather hard to evaluate. However, we see that the potential of the pristine SWCNT film doping has close value to the $[AuCl_4]^-$ complexes reduction (b1), while the potential of treated SWCNTs doping is near the potential of $[AuCl_2]^-$ ions reduction (b3). However, even though the $[AuCl_2]^-$ is a stronger *p*-doping agent, its concentration is relatively small to consider a significant contribution to the overall conductivity of the SWCNT film.

**Table 2.** Measured standard electrode potential and number of transferred electrons of HAuCl$_4$-doped pristine and opened (treated at 350 °C) SWCNT films in comparison to the referential processes of gold chloride anions reduction.

| Discharge process | Standard electrode potential vs. SHE, V | The number of transferred electrons |
|---|---|---|
| Pristine doped | $E^o_{pd} = 1.02 \pm 0.01$* | $n_{pd} = 1.8 \pm 0.4$* |
| Treated at 350 °C doped | $E^o_{td} = 1.12 \pm 0.02$* | $n_{td} = 0.8 \pm 0.1$* |
| $[AuCl_4]^- + 3e^- \rightleftharpoons Au^0 + 4Cl^-$ | $E^o_{b1} = 1.002$ | $n_{b1} = 3$ |
| $[AuCl_2]^- + 1e^- \rightleftharpoons Au^0 + 2Cl^-$ | $E^o_{b3} = 1.154$ | $n_{b3} = 1$ |

* - values calculated using fitting equation $E = E^o + (R \cdot T / n \cdot F) \cdot \ln C_{[AuCl_4]^-}$ that does not account for Cl$^-$ ions concentration.

DFT CALCULATIONS

To reveal the origin of the efficient doping of the heat-treated SWCNTs, we have carried out a theoretical analysis of their electronic properties for separate cases of either only outer or simultaneous outer and inner doping. Earlier density functional theory (DFT) calculations have shown that the *p*-type character of doping is due to the adsorption of AuCl$_4$ species.[32] We can expect other Au-containing species to participate in the doping process. This led us to the systematic investigation of the influence of several AuCl$_x$-based dopants on the electronic structure of the carbon nanotubes. Molecular groups of gold chlorides utilized in calculations were taken from the crystal structure of Au(III)[33] and Au(I)[34] chlorides, while pure Au nanorod was cut from fcc bulk gold. SWCNT of the (10,10) chirality was chosen as a model nanotube to be investigated.

Figure 5 demonstrates analyzed systems of pristine SWCNT (10,10), doped outside, and doped both outside and inside with different dopants located in the SWCNT inner space. Evaluated

Fermi level shifts for each system are presented below the corresponding schemes. The band structures are given in Supporting information (Figure S6).

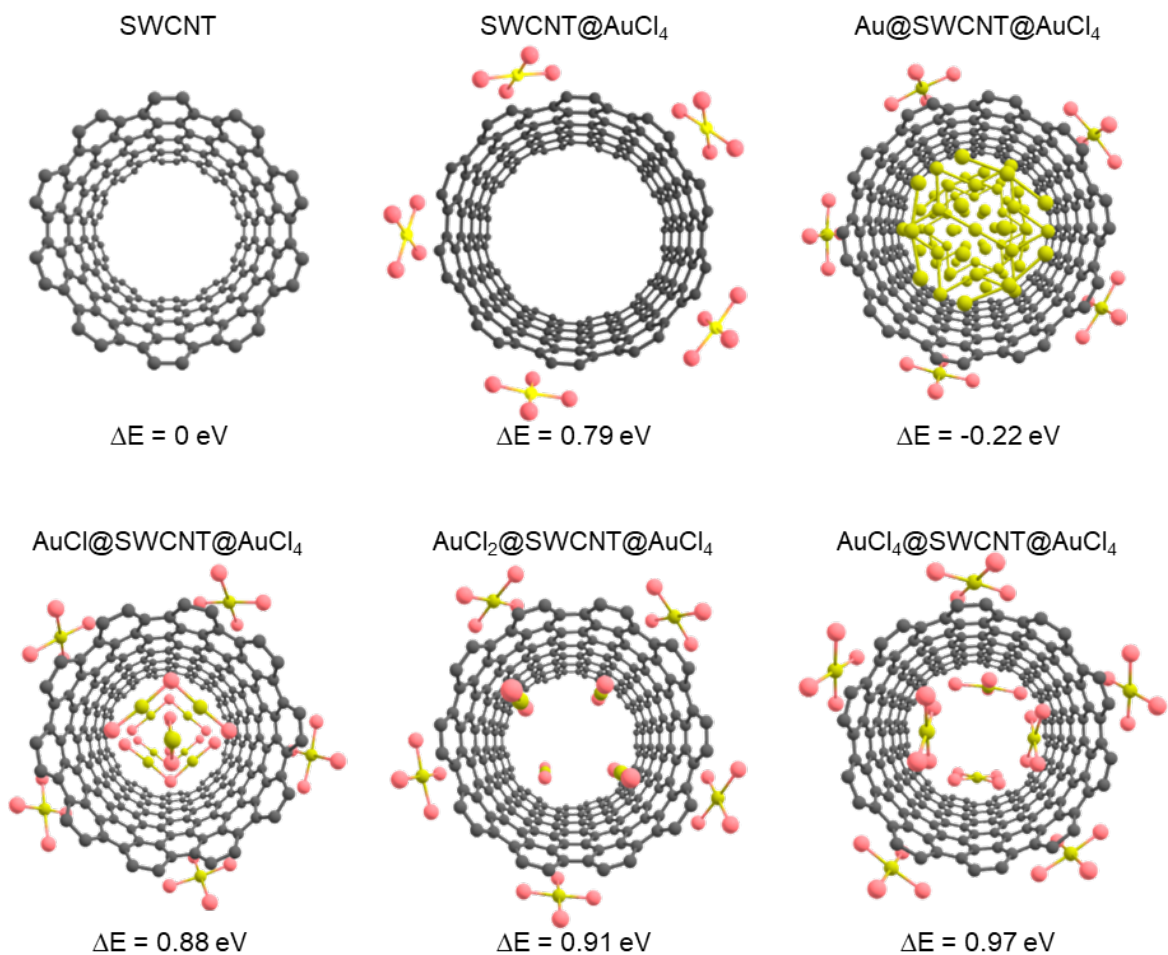

**Figure 5.** Atomic structures of pristine SWCNT(10,10), doped outside, and doped both inside and outside with different $AuCl_x$ dopants. Corresponding Fermi level shift is indicated below each structure. Grey, pink and yellow balls denote carbon, chlorine, and gold atoms, respectively.

Calculations show that SWCNTs doped only from the outside (SWCNT@$AuCl_4$) demonstrate a significant Fermi level downshift for 0.79 eV, characterizing *p*-type doping. Simultaneously, further incorporation of Au nanowires into SWCNTs leads to the charge density transfer from the nanowire to SWCNTs and a 0.22 eV Fermi level general shift, corresponding to *n*-type doping.

In the case of bilateral doping of $AuCl_4$, the largest SWCNT's Fermi level downshift to 0.97 eV is observed. That indicates the more efficient *p*-type doping when compared to the sample doped only from the outer surface, reflecting the efficient doping of opened SWCNTs.

CONCLUSIONS

We developed a novel, simple method of the simultaneous outer surface and inner space SWCNT doping to efficiently modulate the optoelectronic properties of SWCNT films. The proposed method includes a thermal treatment of SWCNT films for the nanotube caps' opening at optimal temperatures of 300 - 400 °C followed by doping with $HAuCl_4$ ethanol solution, thus, providing feasible penetration of dopant solution into the inner SWCNT cavity. The application of the proposed method allowed us to obtain transparent conductive SWCNT films with a record equivalent sheet resistance value of 31 ± 4 Ω/sq. This value was achieved due to significant Fermi-level downshift in case of simultaneous doping from outside and inside of nanotube confirming stronger doping in comparison to pristine tubes with closed ends as proven by DFT-calculations and supported by electrochemical studies.

EXPERIMENTAL SECTION

**SWCNT synthesis.** SWCNTs were synthesized in an aerosol (floating catalyst) chemical vapor deposition (CVD) process in a laminar flow reactor as described elsewhere.[35,36] The process was based on the disproportionation of carbon monoxide on the surface of iron nanoparticles, which are formed in the reactor by means of ferrocene vapor decomposition. Produced SWCNTs were collected onto nitrocellulose membrane filters as films composed of randomly oriented tubes.

The collection time was controlled to collect SWCNT films with the optical transmittance of 70% at the wavelength of 550 nm, which corresponds to the thickness of 37 nm.[37] The mean diameter of collected SWCNTs was calculated from TEM images to be 2.1 nm.

**Thermal treatment and doping.** Pristine SWCNT films were cut with a size of 1 × 1 cm$^2$ and deposited onto quartz substrates using a dry transfer technique.[38] Then, SWCNT films on quartz were heated at the heating rate of 10 °C min$^{-1}$ in a Nabertherm muffle furnace to the desired temperature (up to 500 °C) in the ambient atmosphere. After heat-treatment of the SWCNTs for 20 min, the samples were removed from the furnace and remained at RT to cool down.

The doping procedure was carried out by a drop-casting of a 7 $\mu$l of 15 mM HAuCl$_4$ (HAuCl$_4$·3H$_2$O, ACROS Organics) solution in ethanol (99.5%, ETAX) onto the 1 cm$^2$ SWCNT film at room temperature.

**Characterization.** UV-vis-NIR absorption spectra were measured in the wavelength range of 200 - 3300 nm using the Perkin-Elmer Lambda 1050 spectrophotometer. To ensure the reproducibility of the measurements for each sample, 3 spectra were collected and averaged.

Raman spectroscopy was carried out using a Thermo Fisher Scientific DXRxi Raman Imaging Microscope with a diode-pumped solid-state laser operating at 532 nm. The laser spot diameter on the film surface was around 11 μm. Spectra were averaged using at least 5 measurements. All obtained Raman spectra were normalized to the intensity of the G-mode peak to allow correct comparison.

Sheet resistances of the SWCNT films were measured by the linear four-probe method with Jandel RM3000 Test Unit. The values for each sample were recorded at least 5 times and then averaged. For a quantitative optoelectrical property evaluation of different SWCNT films, we compared their equivalent sheet resistances, $R_{90}$, calculated for the transmittance of 90% at 550 nm.[38]

Transmission electron microscopy was performed using an FEI Tecnai G2 F20 microscope operating at a 100 kV accelerating voltage. The morphology of SWCNT films was analyzed and diameters of SWCNT bundles were estimated based on the sample statistic of around 200 points.

Scanning transmission electron microscope observations were carried out in an aberration-corrected Nion UltraSTEM-100 microscope operated at the 60 keV electron energy. The beam semi-convergence angle was 30 mrad, and the elastically scattered electrons were detected with a medium angle annular dark field (MAADF) detector with a collection semi-angle of 60-200 mrad. MAADF detector was used to improve the contrast of light elements including carbon and thus, to facilitate the accurate determination of SWCNT wall positions in the images. The column vacuum around the sample stage was in the order of $10^{-10}$ mbar during the experiments, suppressing chemical reactions.

**Electrochemical studies.** The open-circuit potential (OCP) measurements were conducted in a three-electrode cell with SWCNTs as a working electrode, Pt counter electrode and KCl saturated AgCl/Cl$^-$ reference electrode (0.197 V *vs*. standard hydrogen electrode, SHE). OCP transients were recorded using BioLogic VMP3 potentiostat/galvanostat, the sampling rate was set to be 0.5 Hz. The OCP measurements were carried out for both pristine and treated SWCNT films, where the treated samples were obtained by treating the pristine SWCNTs at 350 °C according to the procedure described above. The Ohmic drop was estimated to be negligible (*ca.* 0.003 V) and was not considered in OCP measurement.

**DFT calculations.** The calculations were performed by a DFT method within the general gradient approximation (GGA) functional in the Perdew-Burke-Ernzerhof parametrization.[39] We used the projector augmented wave method approximation[40] with the periodic boundary conditions as implemented in the Vienna *Ab-initio* Simulation Package.[41–44] A plane-wave energy cut-off was set to 400 eV. Structural relaxation was performed until the forces acting on each atom became less than $10^{-3}$ eV/Å. Van der Waals interaction was taken into account with Grimme-d3 functional.[45] For the investigation of doping effects, we used 240 carbon atoms per SWCNT(10,10) supercell. Geometry optimization was performed in Gamma point, while for the precise density of states (DOS) and band structure calculations, the $1 \times 1 \times 4$ Monkhorst–Pack scheme was implemented.[46] To avoid spurious interactions between the neighboring unit cells, the translation vector along the non-periodic direction was set to be greater than 15 Å.

ASSOCIATED CONTENT

**Supporting Information**. This material is available free of charge via the Internet at http://pubs.acs.org

Detailed information about thermal treatment, optical and electrical properties of SWCNT films, structural changes of SWCNTs caused by temperature treatment, OCP measurement of pristine and thermally treated SWCNTs, electronic properties calculations, and STEM measurements of doped SWCNTs.

AUTHOR INFORMATION

**Corresponding Author**

*Albert G. Nasibulin: a.nasibulin@skoltech.ru

* Anastasia E. Goldt: a.goldt@skoltech.ru

**Author Contributions**

The manuscript was written through the contributions of all authors. All authors have approved the final version of the manuscript.

**Funding Sources**

The experimental part of the research was supported by the Russian Science Foundation (No. 17-19-01787). Theoretical work was supported by Ministry of Education and Science of the Russian Federation in the framework of Increase Competitiveness Program of NUST "MISiS" (No. K2-2019-016) and Grant of President of Russian Federation for government support of young DSc. (MD-1046.2019.2). This work is supported by the Ministry of Science and Higher Education of the Russian Federation (project no. FZSR-2020-0007 in the framework of the state assignment no. 075-03-2020-097/1). A.P.Ts. acknowledges the EDUFI Fellowship (No. TM-19-11079) from the Finnish National Agency for Education and the Magnus Ehrnrooth Foundation (the Finnish Society of Sciences and Letters) for personal financial support. H.I, J.K., K.M. thank the Austrian Science Fund (FWF) for funding under project no. I3181-N36.

**Notes**

The authors declare no competing financial interest.

ABBREVIATIONS

SWCNTs, single-walled carbon nanotubes; TCFs, transparent and conductive films; OCP, open-circuit potential; ITO, indium-tin-oxide; vHs, van Hove singularities; DFT, density functional theory; RBM, radial breathing mode; TEM, transmission electron microscope; STEM scanning transmission electron microscope; SHE, standard hydrogen electrode; CVD, chemical vapor deposition; MAADF, medium angle annular dark field; GGA, general gradient approximation; DOS, density of states.

(16) Tsapenko, A. P.; Romanov, S. A.; Satco, D. A.; Krasnikov, D. V.; Rajanna, P. M.; Danilson, M.; Volobujeva, O.; Anisimov, A. S.; Goldt, A. E.; Nasibulin, A. G. Aerosol-Assisted Fine-Tuning of Optoelectrical Properties of SWCNT Films. *J. Phys. Chem. Lett.* **2019**, *10* (14), 3961–3965. https://doi.org/10.1021/acs.jpclett.9b01498.

(17) Kim, K. K.; Bae, J. J.; Park, H. K.; Kim, S. M.; Geng, H.-Z.; Park, K. A.; Shin, H.-J.; Yoon, S.-M.; Benayad, A.; Choi, J.-Y.; et al. Fermi Level Engineering of Single-Walled Carbon Nanotubes by AuCl$_3$ Doping. *J. Am. Chem. Soc.* **2008**, *130* (38), 12757–12761. https://doi.org/10.1021/ja8038689.

(18) Kim, K. K.; Yoon, S. M.; Park, H. K.; Shin, H. J.; Kim, S. M.; Bae, J. J.; Cui, Y.; Kim, J. M.; Choi, J. Y.; Lee, Y. H. Doping Strategy of Carbon Nanotubes with Redox Chemistry. *New J. Chem.* **2010**, *34* (10), 2183–2188. https://doi.org/10.1039/c0nj00138d.

(19) Khabushev, E. M.; Krasnikov, D. V.; Zaremba, O. T.; Tsapenko, A. P.; Goldt, A. E.; Nasibulin, A. G. Machine Learning for Tailoring Optoelectronic Properties of Single-Walled Carbon Nanotube Films. *J. Phys. Chem. Lett.* **2019**, 6962–6966. https://doi.org/10.1021/acs.jpclett.9b02777.

(20) Kharlamova, M. V; Yashina, L. V; Volykhov, A. A.; Niu, J. J.; Neudachina, V. S.; Brzhezinskaya, M. M.; Zyubina, T. S.; Belogorokhov, A. I.; Eliseev, A. A. Acceptor Doping of Single-Walled Carbon Nanotubes by Encapsulation of Zinc Halogenides. *Eur. Phys. J. B* **2012**, *85* (1), 34. https://doi.org/10.1140/epjb/e2011-20457-6.

(21) Kharlamova, M. V.; Kramberger, C.; Saito, T.; Shiozawa, H.; Pichler, T. Growth Dynamics of Inner Tubes inside Cobaltocene-Filled Single-Walled Carbon Nanotubes. *Appl. Phys. A Mater. Sci. Process.* **2016**, *122* (8), 1–8. https://doi.org/10.1007/s00339-016-0282-6.23

Supplementary information

# Highly Efficient Bilateral Doping of Single-Walled Carbon Nanotubes

*Anastasia E. Goldt[a], Orysia T. Zaremba[a], Mikhail O. Bulavskiy[a], Fedor S. Fedorov[a],*

*Konstantin V. Larionov[b,c], Alexey P. Tsapenko[d], Zakhar I. Popov[b,e], Pavel Sorokin[b],*

*Anton S. Anisimov[f], Heena Inani[g], Jani Kotakoski[g], Kimmo Mustonen[g], Albert G. Nasibulin[a,h]\**

[a]Skolkovo Institute of Science and Technology, 3 Nobel Street, 121205 Moscow, Russia

[b] National University of Science and Technology "MISiS", Leninsky prospect 4, 119049 Moscow, Russian Federation

[c]Moscow Institute of Physics and Technology, Institutskiy lane 9, Dolgoprudny, 141700 Moscow region, Russian Federation

[d]Aalto University, Department of Applied Physics, Puumiehenkuja 2, 00076 Espoo, Finland

[e]Emanuel Institute of Biochemical Physics RAS, 119334 Moscow, Russian Federation

[f]Canatu Ltd., Konalankuja 5, 00390 Helsinki, Finland

[g]University of Vienna, Faculty of Physics, Boltzmanngasse 5, 1090 Vienna, Austria

[h]Aalto University School of Chemical Engineering, Kemistintie 1, 16100 Espoo, Finland

1. **Thermal Treatment Optimization**

To optimize our method for efficient SWCNT doping utilizing their inner space for dopant incorporation, we evaluated optoelectronic and structural changes, induced by SWCNT thermal treatment in the range of 100 – 500 °C. For these purposes, we used UV-vis-NIR and Raman spectroscopies, TEM imaging, and four-probe sheet resistance measurements.

**1.1. Changes in Optical and Electrical Properties Induced by Thermal Treatment**

UV-vis-NIR absorption spectra were measured to analyze the electronic states of SWCNTs before and after the thermal treatment (Figure S1a). Spectra usually clearly display the excitonic transitions between van Hove singularities in semiconducting ($E_{11}$ and $E_{22}$) and metallic ($M_{11}$) SWCNTs. It should be noted that SWCNT films are environmentally doped by adsorbed gaseous species. In order to clearly represent the processes occurring during the thermal treatment in the temperature range examined, we plotted the absorbance values at the wavelengths of 2378, 1310, and 894 nm, which corresponded to the mean value of $E_{11}$, $E_{22}$, and $M_{11}$ transitions (Figure S1b). $S_{11}$ peak intensity exhibits significant monotonous growth with the temperature increase up to 350 °C, the intensity of $S_{22}$ peak is characterized by the same behavior pattern, although with a smaller growth rate, and the $M_{11}$ peak remains nearly unaltered up to 350 °C. When the treatment temperatures surpass 350 °C, all peaks start to decrease remarkably. The initial growth of $S_{11}$ and $S_{22}$ peak intensities with temperature up to 350 °C can be attributed to the desorption of oxygen, which is known to be adsorbed on the SWCNT surface from the air leading to the *p*-doping effect.[S1] Rapid drop of all peak intensities at higher temperatures (400 °C and above) indicates the oxidation of SWCNTs, which starts from end-caps since they are the most reactive parts of nanotubes.[S2]

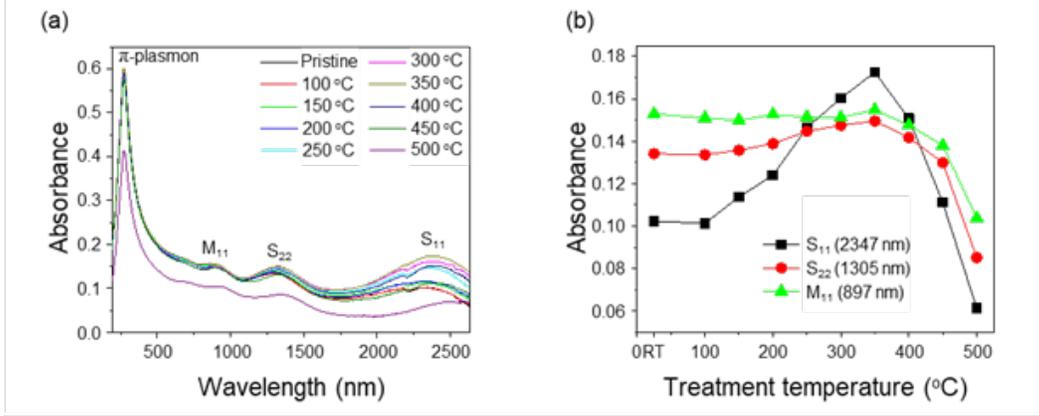

**Fig. S1** UV-vis-NIR absorption spectra of the SWCNT films treated at different temperatures (a). $S_{11}$ (2378 nm), $S_{22}$ (1310 nm), and $M_{11}$ (894 nm) peaks as a function of the treatment temperature (b).

Since transmittance ($T$) and equivalent sheet resistance ($R_{90}$) are quality defining parameters for transparent conductors, we monitored their variations with the change in the treatment temperature. Figure S2 represents relative changes in $T$ and $R_{90}$ in comparison to the pristine sample, as functions of the treatment temperature. Equivalent sheet resistance was calculated as:

$$R_{90} = R_s \cdot \frac{A}{\log_{10}\left(\frac{10}{9}\right)}, \qquad (1)$$

where A and $R_s$ are the absorbance (at 550 nm) and sheet resistance of the films.[S3]

The transmittance (Figure S2) remains almost unchanged when SWCNT films are treated at temperatures up to 350 °C, although at higher temperatures the transmittance increases markedly. The equivalent sheet resistance exhibits continuous growth with the increase of the treatment temperature, which can be explained by the desorption of oxygen from the SWCNT surface and the oxidation of SWCNTs at temperatures higher than 350 °C.

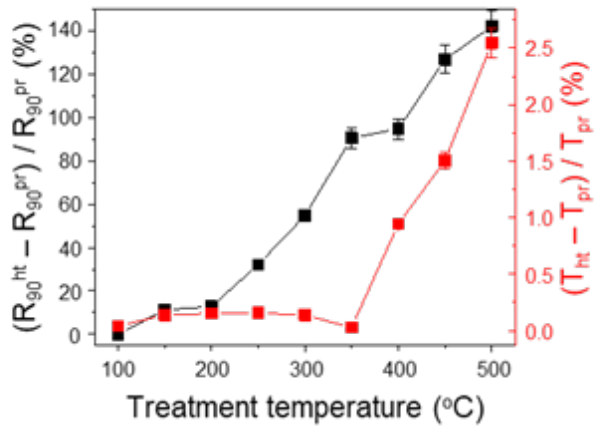

**Fig. S2** Relative changes of transmittance and equivalent sheet resistance of SWCNT films *versus* the treatment temperature, ht signifies heat treated.

### 1.2. Structural Changes Caused by Thermal Treatment

The quality of SWCNTs was estimated as the ratio of intensities $I_G/I_D$ calculated using Raman spectroscopy data. The thermal treatment of the SWCNT films up to 300 °C did not lead to significant changes in the $I_G/I_D$ ratio. Further increase in the temperature caused a slight decrease in the $I_G/I_D$ ratio, which might be attributed to the introduction of defects in the SWCNTs due to the oxidation process in the air (Figure S3a). These results are in accordance with the changes in the intensity distribution in the RBM frequency range with increasing treatment temperature (Figure S3a). The RBM peaks remain unaltered up to 300 °C, and a further increase in the temperature leads to the disappearance of RBM peaks corresponding to smaller diameters, which are more reactive.[S4]

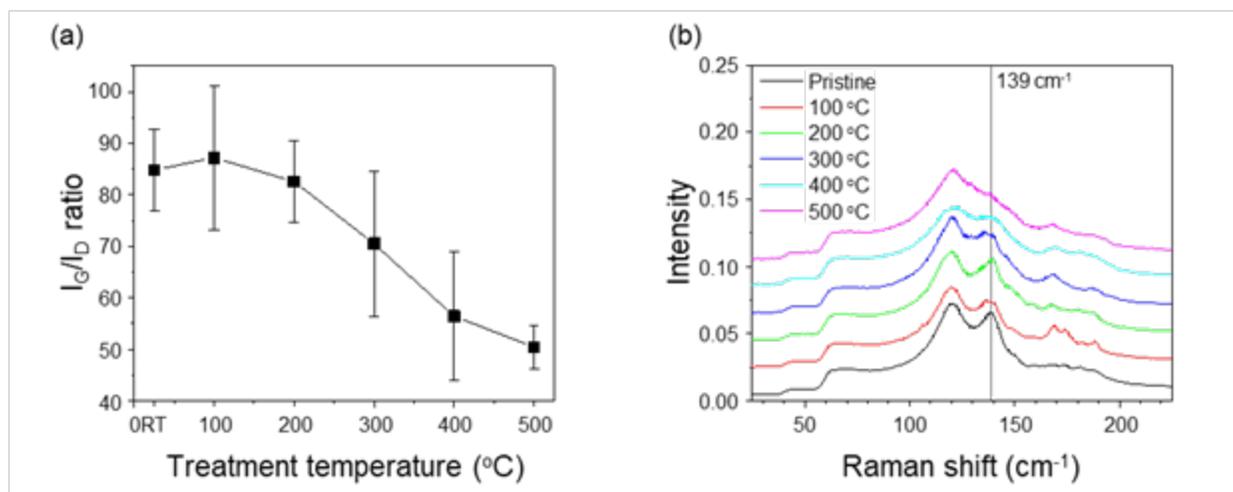

**Fig. S3** Raman spectra alterations with thermal treatment of SWCNT films at different temperatures. Evolution of the $I_G/I_D$ intensity ratio (a) and RBM-mode after treatment at different temperatures (b).

Figure S4 shows the typical TEM images of pristine SWCNT films and thermally treated at 300 and 400 °C. The morphology of the SWCNTs after the thermal treatment remained similar when compared to the pristine SWCNTs. We have plotted the diameter of bundles diameter, and it follows the Gaussian distribution. The calculation of mean diameters of bundles at different temperatures revealed that thermal treatment above 300 °C leads to a small increase in the average thickness of bundles from 7 (300 °C) to 10 (400 °C) nm due to an increase in the thermal energy during the treatment process.

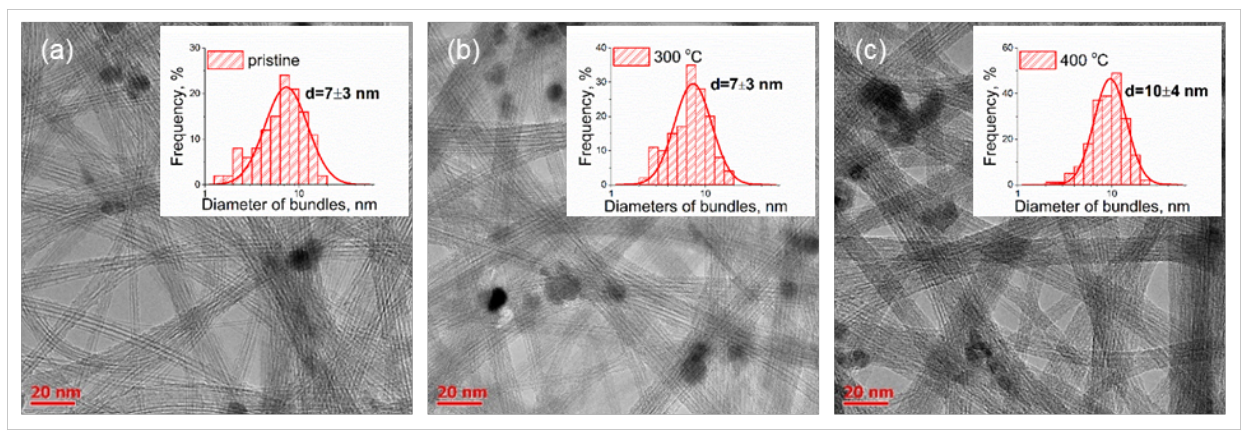

**Fig. S4** Transmission electron microscopy images of SWCNTs films and corresponding bundle diameter distributions (insets) for the pristine film (a), after thermal treatment at 300 °C (b) and 400 °C (c).

## 2. HAuCl₄ Doping after Thermal Treatment of SWCNTs

To investigate the efficiency of the doping effect after thermal treatment and the formation of gold nanowires in the inner space of SWCNTs, we used Raman and XPS spectroscopies, STEM imaging, OCP measurements, and band structure calculations.

### 2.1. Raman Measurements of Doped SWCNTs

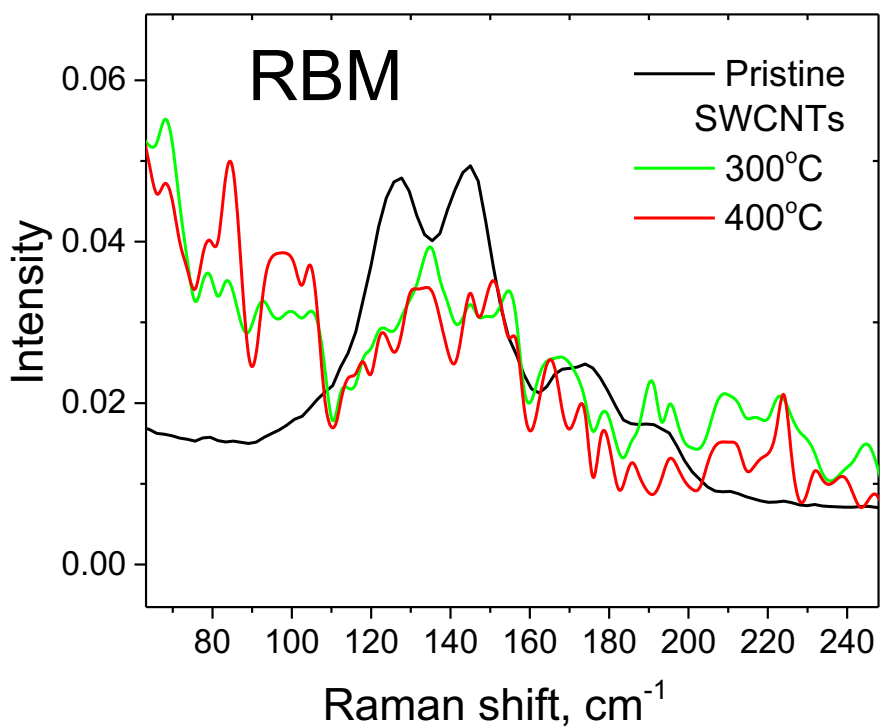

**Fig. S5** RBM Raman spectra of HAuCl₄-doped SWCNTs: pristine and after thermal treatment at 300 and 400 °C.

## 2.2. STEM Measurements of Doped SWCNTs

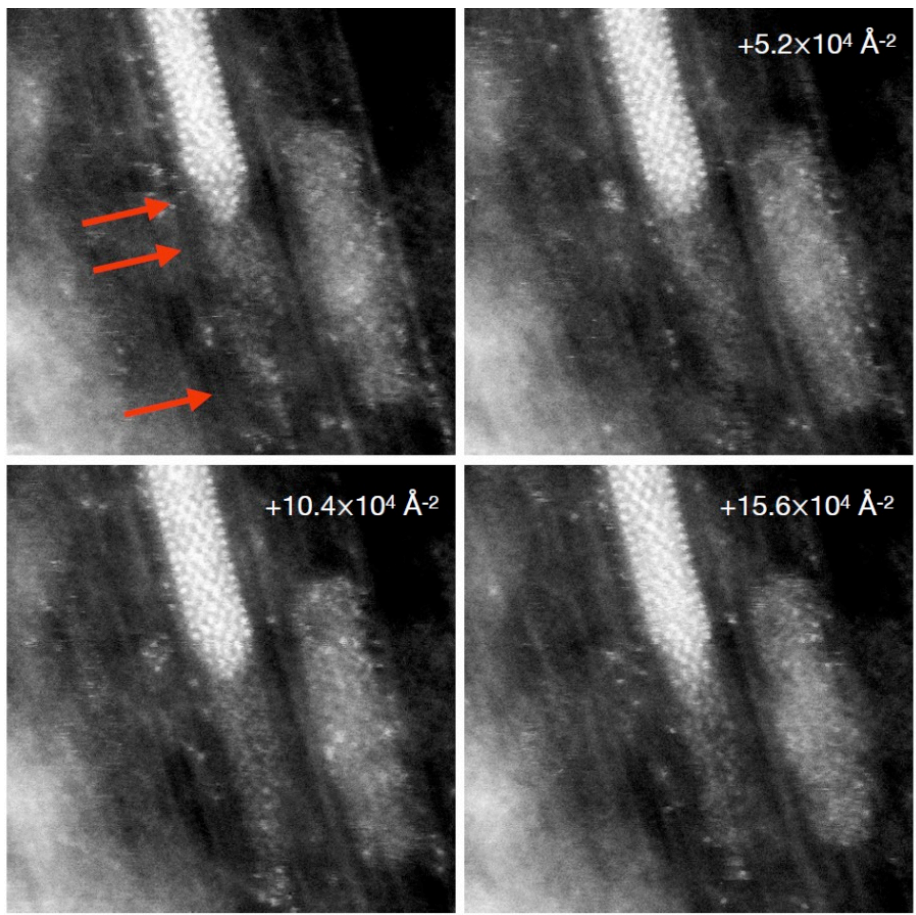

**Fig. S6** *In situ* diffusion of gold along the nanotube axis, the position pointed out by the arrows (see also video). The diffusion process is driven by the scanning electron probe, the dose equivalent per each acquired frame is 5.2 x $10^4$ e$^-$/Å$^2$ at 25 pA probe current. The field of view is 8 × 8 nm$^2$.

## 2.3. OCP Measurement of Doped SWCNTs with and without Thermal Treatment

Figure S7 shows the OCP transient curves recorded for all probed SWCNT films: pristine (Figure S7a) and treated at 350 °C (Figure S7b) SWCNT films, immersed into HAuCl$_4$ ethanol solutions of concentrations varied from 7.5 to 60 mM.

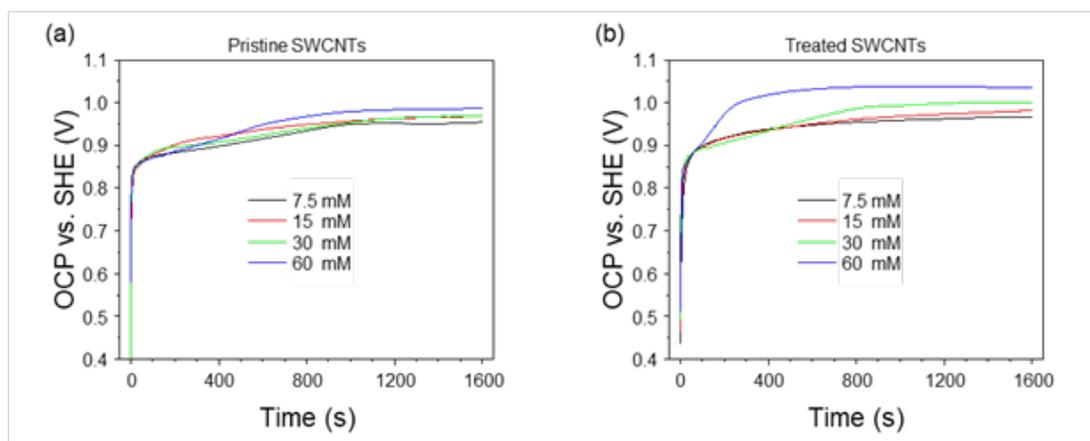

**Fig. S7** OCP transient for pristine (a) and thermally treated SWCNTs (b) recorded in solutions of 7.5, 15, 30 and 60 mM HAuCl$_4$ concentrations.

### 2.4. XPS Measurements of Doped SWCNTs

The high-resolution XPS spectra of Au 4f and Cl 2p of 15 mM HAuCl$_4$-doped SWCNTs are depicted in Figure S8. By deconvolving the asymmetric Au 4f doublets, we can observe four components such as the reduced Au$^0$ clusters (Au$^0$ 4f$_{5/2}$ and Au$^0$ 4f$_{7/2}$) and the Au$^{3+}$ ions (Au$^{3+}$ 4f$_{5/2}$ and Au$^{3+}$ 4f$_{7/2}$), as shown in Figure S8a. The higher peaks of Au$^{3+}$ indicate the substantial amount of Au$^{3+}$ that remains on nanotubes. This is confirmed by the Cl 2p core peak, which is attributed to metal chloride (Figure S8b).

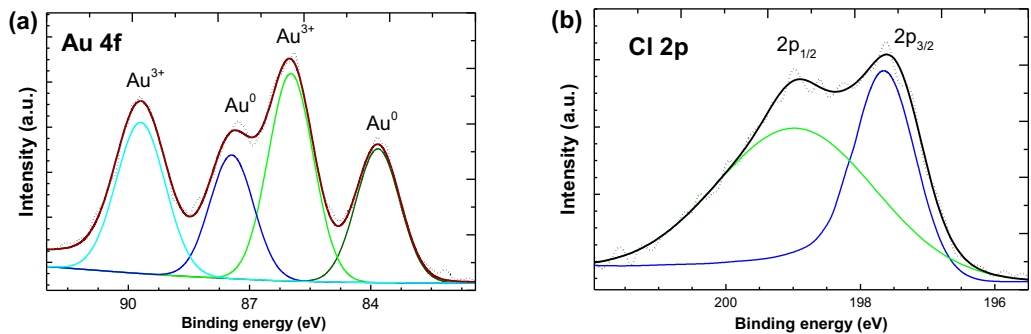

**Fig. S8** XPS analysis of fitted (a) Au 4f and (b) Cl 2p core peaks.

## 2.5. Electronic Properties Calculations

Figure S9 shows band structures corresponding to pristine SWCNT (10,10), the same tube doped with $AuCl_4$ from the outside, and several additionally doped SWCNTs from the inside with different $AuCl_x$ dopants.

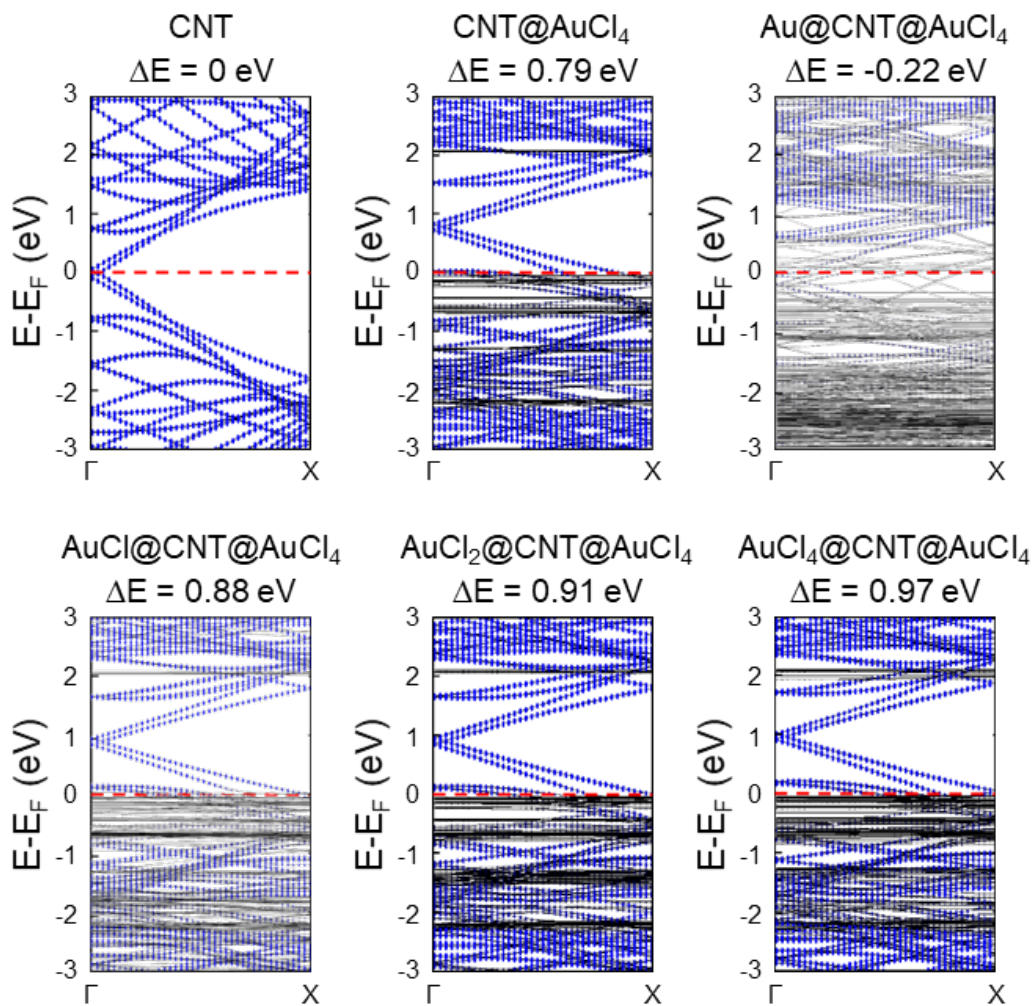

**Fig. S9** Band structures of pristine SWCNT (10,10), doped with $AuCl_4$ from the outside, and doped from both the inside and outside with different $AuCl_x$ dopants as shown in Figure 5. Fermi level is indicated with a red dashed line, the blue dots correspond to contributions from carbon atoms.